\documentclass[%
reprint,
superscriptaddress,
aps,
pra,
twocolumns,
]{revtex4-2}
\usepackage[left=1.5cm, right=1.5cm, top=1.785cm, bottom=2.0cm]{geometry}
\usepackage{balance}
\usepackage{times,mathptmx}
\usepackage{graphicx} 
\usepackage{lastpage}
\usepackage{caption}
\usepackage{float}
\usepackage{fancyhdr}
\usepackage{fnpos}
\usepackage[english]{babel}
\usepackage{array}
\usepackage{droidsans}
\usepackage{charter}
\usepackage[T1]{fontenc}
\usepackage[usenames,dvipsnames]{xcolor}
\usepackage{setspace}
\usepackage[compact]{titlesec}
\usepackage{hyperref}
\usepackage[utf8]{inputenc} 
\usepackage{tabularx}
\usepackage{color}				
\usepackage{amsmath}
\usepackage{amsthm}
\usepackage{amsfonts}
\usepackage{mathtools}
\usepackage{lmodern} 			
\usepackage{lipsum}
\usepackage{adjustbox}
\usepackage{pgfplots}			
\usepackage{tikz}					
\usetikzlibrary{shapes}
\usetikzlibrary{calc}
\usetikzlibrary{positioning}
\usepackage{tikzscale}		
\usepackage{latexsym} 		
\usepackage{subcaption}				
\usepackage{listings}
\usepackage{marginnote}
\usepackage[normalem]{ulem}
\usepackage{etoolbox}
\usepackage{appendix}
\usepackage{soul}
\usepackage{dcolumn}				
\usepackage{bm}						
\usepackage[mathlines]{lineno}		

\definecolor{naranja_1}{rgb}{0.70,0.25,0}
\definecolor{gris-azul}{rgb}{0.27,0.33,0.415}  
\definecolor{azul-gris}{rgb}{0.192,0.278,0.4}  
\definecolor{verde-chulo}{rgb}{0.25,0.5,0.5}  
\definecolor{verde-chulo2}{rgb}{0,0.5,0.4} 
\definecolor{beige}{rgb}{0.96, 0.96, 0.86}

\definecolor{orcidlogocol}{HTML}{A6CE39}

\definecolor{color1}{rgb}{1, 0, 0}%
\definecolor{color2}{rgb}{0, 0, 1}%
\definecolor{color3}{rgb}{0, 1, 0}%
\definecolor{color4}{rgb}{1, 0, 1}%
\definecolor{color5}{rgb}{0, 1, 1}%

\definecolor{mycolor1}{rgb}{0.00000,0.44700,0.74100}%
\definecolor{mycolor2}{rgb}{0.85000,0.32500,0.09800}%
\definecolor{mycolor3}{rgb}{0.46600,0.67400,0.18800}%
\definecolor{mycolor4}{rgb}{0.49400,0.18400,0.55600}%
\definecolor{mycolor5}{rgb}{0.92900,0.69400,0.12500}%
\definecolor{mycolor6}{rgb}{0.30100,0.74500,0.93300}%
\definecolor{mycolor7}{rgb}{0.63500,0.07800,0.18400}%

\pgfplotsset{
	colormap={parula}{
		rgb255=(53,42,135)
		rgb255=(15,92,221)
		rgb255=(18,125,216)
		rgb255=(7,156,207)
		rgb255=(21,177,180)
		rgb255=(89,189,140)
		rgb255=(165,190,107)
		rgb255=(225,185,82)
		rgb255=(252,206,46)
		rgb255=(249,251,14)
	},
}

\pgfplotsset{compat=1.5}
\pgfplotsset{minor grid style={dashed}}

\hypersetup{
	colorlinks = true,
	linkcolor = azul-gris,
	citecolor = naranja_1,
	urlcolor = orange,
	runcolor = black,
	breaklinks = true,
}

\newcommand{\secref}[1]{\hyperref[#1]{Section \ref*{#1}}}		
\newcommand{\appref}[1]{\hyperref[#1]{Appendix \ref*{#1}}}		
\newcommand{\chref}[1]{\hyperref[#1]{Chapter \ref*{#1}}}		
\newcommand{\figref}[1]{\hyperref[#1]{Figure \ref*{#1}}}		
\newcommand{\sfigref}[2]{\hyperref[#1]{Figure \ref*{#1}#2}}		
\newcommand{\sfigrefN}[2]{\hyperref[#1]{\ref*{#1}#2}}			
\newcommand{\tabref}[1]{\hyperref[#1]{Table \ref*{#1}}}			

\pgfkeys{/tikz/savenumber/.code 2 args={\global\edef#1{#2}}}

\newcolumntype{L}[1]{>{\raggedright\arraybackslash}p{#1}}
\newcolumntype{C}[1]{>{\centering\arraybackslash}p{#1}}
\newcolumntype{R}[1]{>{\raggedleft\arraybackslash}p{#1}}

\usepackage{academicons}
\usepackage{epstopdf}

\begin{document}	
	\title{Multi-scale analysis of radio-frequency performance of 2D-material based field-effect transistors}
	
	\author{A. Toral-Lopez}
	\email{atoral@ugr.es}
	\affiliation{Departamento de Electrónica y Tecnología de Computadores, Facultad de Ciencias, Universidad de Granada (Spain)}%
	
	\author{F. Pasadas}
	\affiliation{Departament d'Enginyeria Electrònica,  Universitat Autònoma de Barcelona, 08193, Bellaterra (Spain)}%
	
	\author{E.G. Marin}
	\affiliation{Departamento de Electrónica y Tecnología de Computadores, Facultad de Ciencias, Universidad de Granada (Spain)}%
	
	\author{A. Medina-Rull}
	\affiliation{Departamento de Electrónica y Tecnología de Computadores, Facultad de Ciencias, Universidad de Granada (Spain)}%
	
	\author{J. M. Gonzalez-Medina}
	\affiliation{Global TCAD Solutions GmbH., Bösendorferstraße 1/12, 1010, Vienna (Austria)}%
	
	\author{F.G. Ruiz}
	\affiliation{Departamento de Electrónica y Tecnología de Computadores, Facultad de Ciencias, Universidad de Granada (Spain)}%
	\affiliation{Pervasive Electronics Advanced Research Laboratory, CITIC, Universidad de Granada, 18071, Granada (Spain)}
	
	\author{D. Jimenez}
	\affiliation{Departament d'Enginyeria Electrònica,  Universitat Autònoma de Barcelona, 08193, Bellaterra (Spain)}%
	
	\author{A. Godoy}
	\email{agodoy@ugr.es}
	\affiliation{Departamento de Electrónica y Tecnología de Computadores, Facultad de Ciencias, Universidad de Granada (Spain)}%
	\affiliation{Pervasive Electronics Advanced Research Laboratory, CITIC, Universidad de Granada, 18071, Granada (Spain)}
	
	\begin{abstract}		
		Two-dimensional materials (2DMs) are a promising alternative to complement and upgrade high-frequency electronics. However, in order to boost their adoption, the availability of numerical tools and physically-based models able to support the experimental activities and to provide them with useful guidelines becomes essential. In this context, we propose a theoretical approach that combines numerical simulations and small-signal modeling to analyze 2DM-based FETs for radio-frequency applications. This multi-scale scheme takes into account non-idealities, such as interface traps, carrier velocity saturation, or short channel effects, by  means of self-consistent physics-based numerical calculations that later feed the circuit level via a small-signal model based on the dynamic intrinsic capacitances of the device. At the  circuit stage, the possibilities range from the evaluation of the performance of a single device to the design of complex circuits combining multiple transistors. In this work, we validate our scheme against experimental results and exemplify its use and capability assessing the impact of the channel scaling on the performance of MoS$_2$-based FETs targeting RF applications.
	\end{abstract}
	
	\maketitle

\section{Introduction}\label{sec:intro}
The emergence  of two-dimensional (2D) crystals has raised the expectations for radio-frequency (RF) electronics to move into the THz range. This  territory is especially suitable for graphene thanks to its high mobility and saturation velocity which provide an excellent basis for RF operation. In fact, graphene field-effect transistors (FETs) with several hundreds of GHz cut-off frequencies ($f_{\rm T}$) have already been demonstrated \cite{Cheng2012,Liao2010}. Their maximum-oscillation frequency ($f_{\rm max}$), however, is still low, $\leq 200$~GHz \cite{Wu2016,Feng2014,Guo2013},  due to the high output conductance produced by the absence of a bandgap \cite{Schwierz2013}. Other 2D materials (2DMs) are postulated as candidates to overcome this problem, being MoS$_2$, with a bandgap sizable with the number of stacked layers,  one of the most appealing and studied alternatives \cite{Krasnozhon2014, Gao2018, Chang2015, Sanne2015, Sanne2017, Cheng2014, Belete2018}. 

The possibility to transfer 2DMs  in almost any substrate expands the range of applications with respect to conventional RF technologies. Indeed, measurements of  MoS$_2$ devices fabricated on flexible substrates show values of $f_{\rm T}=13.5$GHz and $f_{\rm max}=10.5$GHz \cite{Cheng2014}, not that far from  RF performances achieved with the same material on rigid oxides, $f_{\rm T}=42$GHz and $f_{\rm max}=50$GHz \cite{Cheng2014}. {Better RF performance would be achievable as further improvements come in MoS$_2$ and other promising 2DMs such as PdSe$_2$\cite{Bartolomeo2019} or PtSe$_2$ \cite{Urban2020}. In this regard, while 2D semiconductors are constrained by a lower $f_{\rm T}$ than graphene (due to the superior carrier mobility and saturation velocity of the latter), they are not limited by the gapless band-structure that impacts the current saturation in GFETs leading to a reduced voltage and power gain \cite{Schwierz2013}. Indeed, GFETs have behaved rather poor in terms of $f_{\rm max}$\cite{Schwierz2013}, limiting the highest operating frequency of power amplifiers. It is in this field where more research is required to shed light on the potential of MoS$_2$ and other TMDs to develop improved RF applications \cite{Schwierz2015}.}

In this arena, with the 2D RF technology readiness in its early infancy, the computational tools able to support and rationalize the experimental activities are more essential than ever so as to: {\it{i}}) interpret the experimental measurements, {\it{ii}}) gain understanding to improve the capabilities of the devices, {\it{iii}}) assess their performance and benchmark them against conventional technologies, {\it{iv}}) explore new device designs, and {\it{v}}) foresee the RF limits of novel technologies. 

In this context, this work presents a multi-scale scheme \cite{Marin2018b} that combines numerical simulations with a small-signal model to analyze the RF performance of 2DM-based FETs. The fully detailed behavior of the device in the static operation regime obtained from the numerical simulator is combined with a small-signal model specifically developed for 2DM-based FETs \cite{Pasadas2019}. Thus, we are able to extend the analysis from a device to a circuit level. To this purpose, we first introduce the theoretical basis of the calculations. Then, the multi-scale method is validated  against the dc and high-frequency measurements of MoS$_2$ transistors reported in \cite{Krasnozhon2014}. Finally, $f_{\rm T}$ and $f_{\rm max}$ are evaluated for different gate lengths to analyze the RF potential performance of MoS$_2$ transistors, and then the main conclusions of the work are outlined.

\section{Methods}\label{sec:methods}

The multi-scale approach proposed in this work combines {\it{i}}) the numerical self-consistent solution of the coupled Poisson and continuity equations, 
including interface traps, carrier velocity saturation and short-channel effects;  and {\it{ii}}) the small-signal compact model that allows for the analysis of multiple interconnected devices in arbitrary designs, and their linear response to complex dynamical inputs.  

\begin{figure}[th]
	\centering
	\includegraphics[width=\linewidth]{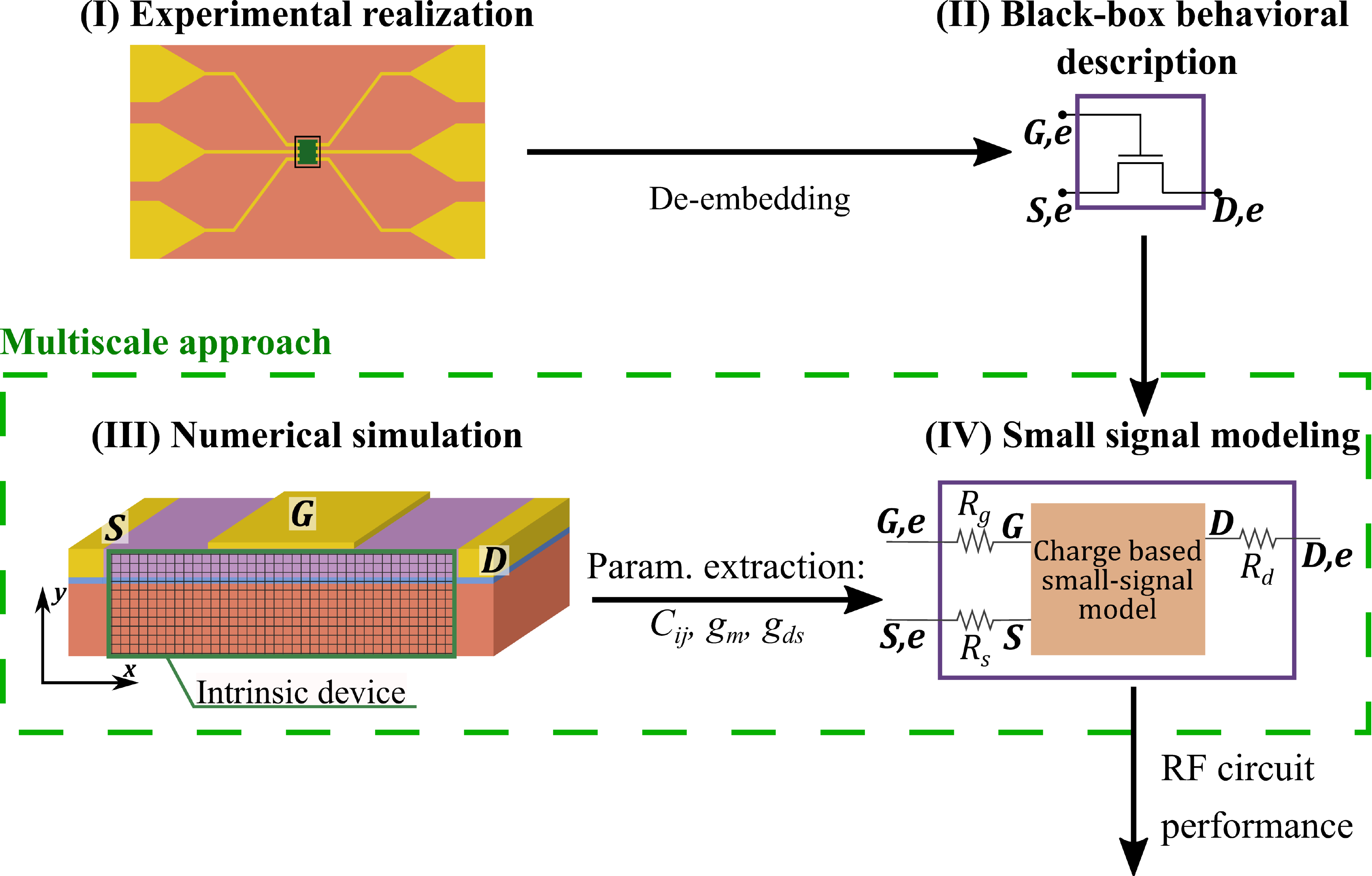}
	\caption{The multi-scale approach consists on combining a small-signal equivalent circuit and a numerical simulator to describe the behavior of a 2DM-based FETs. The small-signal elements are extracted from the numerical solver and fed into the small-signal equivalent circuit. Importantly, the metal-2DM contact resistances as well as the gate resistance are also included.}
	\label{fig:devicemodellingsim}
\end{figure}

With the purpose of tighting the approach to the state-of-the-art, and make realistic prospects for that technology, the studied architecture can be taken from an experimental setup of a particular 2DM-based device. Then, a self-consistent numerical simulation in the static regime is carried out, providing detailed information of the main electrostatic and transport physical magnitudes governing the transistor response, including the intrinsic device and the parasitic elements involved to contact it.  Next, the intrinsic magnitudes can be extracted and fed into the small-signal model, so to have a distilled description of the device physics. The extrinsic and parasitic elements are thus  isolated, and can be later re-introduced in the small-signal model, becoming external, technology-dependent parameters to the core model. That approach enables to project the RF performance of the devices and materials regardless the current technology constraints and provides a flexible and at the same time robust description. The overall picture of the proposed procedure is shown in \figref{fig:devicemodellingsim}. The stages are described in more detail hereunder.

\subsection{Self-consistent numerical simulator (device level)}

\subsubsection{Static operation: electrostatics and carrier transport }

First, the self-consistent solution of the Poisson and continuity equations is obtained for a given material and device architecture, assuming a drift-diffusion transport regime, which constitutes an accurate description for the state-of-the-art device dimensions.  The electrostatic potential, the carrier density profiles, the quasi-Fermi levels for electron and holes, mapped as a function of the spatial position, and the drain-to-source current ($I_{\rm DS}$) are obtained, for every combination of terminal biases.

The impact of non-idealities is also included in the simulations. In particular, three remarkable effects  are considered: interface traps, electric field dependent mobility and access and contact resistances. {The impact of the gate electric field on the mobility  might be also of relevance as it has been studied in \cite{Bartolomeo2017} and \cite{Feijoo2019} but requires a detailed analysis of the scattering processes in MoS$_2$}. For the former, an arbitrary energetic profile for either donors or acceptors traps can be defined to evaluate the surface charge density associated with a certain interface as a function of the electrostatic potential and Fermi level (see S4 at SI). As for the mobility, we consider the electric field induced degradation (see S5 at SI), following the expression proposed in \cite{Feijoo_2016}.

When analyzing a device, the numerical modeling described above is applied to the complete structure, as depicted in \figref{fig:devicemodellingsim}, which comprises channel plus access regions and contact resistances. This enables a full self-consistent solution of the electrostatic and transport dependencies in the device. Then, the contact resistances, that are bias independent, are removed with the aim of simulating the intrinsic device and extract its small-signal parameters as explained in the following.

\subsubsection{ Dynamic operation: terminal charges and intrinsic capacitance scheme}

To compute the dynamic operation of the device, the charge associated with each terminal is evaluated as a function of the different biases in order to subsequently determine their intrinsic capacitances \cite{Pasadas2016}. In particular, the charges associated with the gate, drain and source terminals are calculated following the Ward-Dutton charge partition scheme, which ensures the charge conservation \cite{Ward1978} (see S7-S9 at SI).

Then, we compute the reference-independent dynamic description of a three-terminal device by calculating the intrinsic capacitances, $C_{ij}$, as the charge derivative at terminal $i$ with respect to a varying voltage applied to terminal $j$, assuming that the bias at any other terminal remains constant:
\begin{equation}
	C_{ij} = \left\{
	\begin{array}{ll}
		\frac{\partial Q_{i}}{\partial V_{j}}  &  i = j \\
		-\frac{\partial Q_{i}}{\partial V_{j}} &  i \neq j
	\end{array}
	\right. , \,\,\, i,j = \rm g,d,s
	\label{eq:capcalc}
\end{equation}
where the subscripts g, d and s stand for gate, drain and source, respectively \cite{Pasadas2019}.

\subsection{Small-signal model (circuit level)}

\begin{figure}[th]
	\centering
	\includegraphics[width=\linewidth]{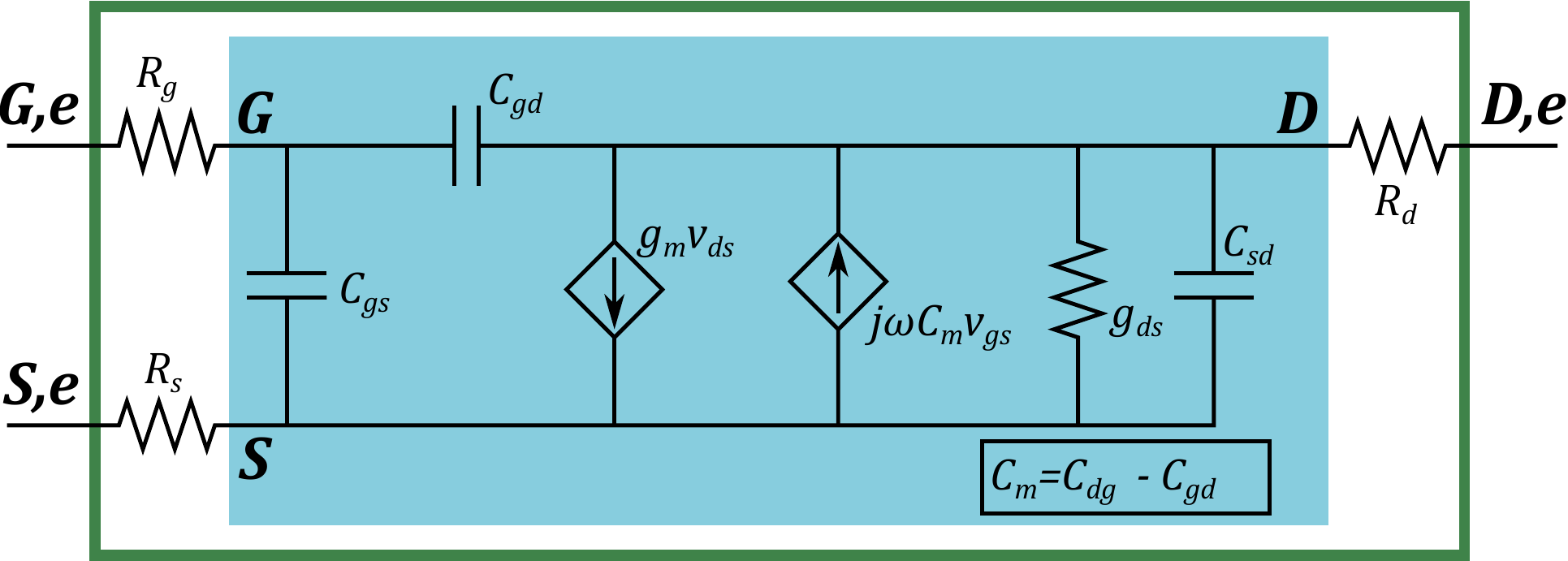}
	\caption{Small-signal equivalent circuit suited to 2DM-based FETs \cite{Pasadas2017}. The equivalent circuit of the intrinsic device is framed in blue. The small-signal elements are: $g_{\rm m} =\partial I_{\rm DS}/\partial V_{\rm GS}$ transconductance; $g_{\rm ds}=\partial I_{\rm DS}/\partial V_{\rm DS}$ output conductance; and $C_{\rm gs}$, $C_{\rm gd}$, $C_{\rm sd}$, and $C_{\rm dg}$ intrinsic capacitances. $R_{\rm g}$ is the gate resistance and $R_{\rm d}$ and $R_{\rm s}$ account for the contact resistances of the drain and source, respectively. They connect the intrinsic (noted {\it{G}}, {\it{D}} and {\it{S}}) and extrinsic  ({\it{G,e}}, {\it{D,e}} and {\it{S,e}}) gate, drain and source terminals.}
	\label{fig:devicemodellingssig}
\end{figure}

The small-signal equivalent circuit of a 2DM-based FET is shown in \figref{fig:devicemodellingssig} \cite{Pasadas2017}. It allows to treat the  current and charge variations due to a time-varying input signal in terms of linear circuit elements, i.e. conductance, and capacitance elements (see SI). Two main features must be highlighted: {{\it{i}}) the metal contact resistances are included, an aspect of utmost importance when dealing with low dimensional FETs \cite{Jena2014} and {\it{ii}}) the small-signal model guarantees charge conservation and takes into account non-reciprocal capacitances. The latter is not considered in Meyer \cite{Meyer} and Meyer-like capacitance models making them inaccurate when trying to predict the RF performance of GFETs, where the impact of transcapacitances is critical. Indeed, the analysis of the intrinsic capacitances of a MoS$_2$ FET addressed in \cite{Pasadas2019} shows that reciprocity between C$_{\rm gd}$ and C$_{\rm dg}$ cannot be assumed for all transistor operation regions.} Such an equivalent circuit combined with our detailed self-consistent physics-based simulator allows the accurate electrical simulation of 2DM-based FETs for linear RF applications \cite{ToralLopez2019}.

Considering the equivalent circuit shown in \figref{fig:devicemodellingssig} as a two-port network connected in a common-source configuration allows the assessment of two main figures of merit for RF devices: the cut-off frequency, $f_{\rm T}$, and the maximum oscillation frequency, $f_{\rm max}$, evaluated from the current and power gain, respectively. 

\section{Results and Discussion}

Before projecting the potential performance of MoS$_2$ FETs for RF applications, the multi-scale scheme is validated against the experimental data reported in \cite{Krasnozhon2014}. The device is a monolayer MoS$_2$ FET with 30nm-thick top gate oxide (HfO$_2$) and a 270nm-thick substrate (SiO$_2$). The total length of the device is $L_{\rm ch}=340$~nm, with a 240nm-long channel plus two access regions 50nm-long each. The measured contact resistance and electron mobility are 2~k$\Omega\cdot\mu$m and 85~cm$^2$/Vs, respectively.

\begin{figure}[th]
	\centering
	\includegraphics[width=\linewidth]{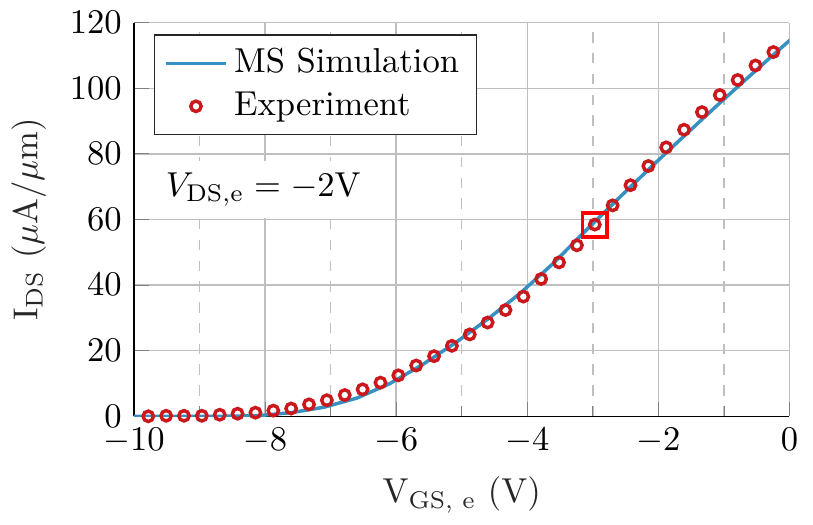}
	\caption{Experimental data of a MoS$_2$ FET reported in \cite{Krasnozhon2014} (symbols) and results from the numerical solver (line). The operating point for RF experimental measurements is indicated by a square.}
	\label{fig:ajusteidsvfg}
\end{figure}

With the intention of emulating the effect of the contact resistances ($R_{\rm c}$) in the numerical simulations, two doped regions are added at both edges of the semiconductor layer and its doping is modified to adjust $R_{\rm c}=R_{\rm s}=R_{\rm d}$ to the reported experimental value (2~k$\Omega\cdot\mu$m). Interface traps {are a relevant magnitude in MoS$_2$ devices, as it is discussed in \cite{Dagan2020,Kim2017}, although more effort is still required to achieve a deeper and more comprehensive understanding. However, as the detailed distribution of interface traps is not the main goal of this work, we have considered a constant energetic profile in both interfaces of the channel that provides the best agreement with the experimental data \cite{Takenaka2016}. In particular, we set two energetic profiles: }({\it{i}}) at the top gate insulator interface, a constant energetic profile of donors traps with $D_{\rm it}=10^{12}$cm$^{-2}$eV$^{-1}$ is considered from mid-gap up to $0.57$~eV towards the conduction band edge, and {\it{ii}}) at the substrate interface, a constant energetic profile also of donor traps from midgap to the conduction band edge (i.e. $0.9$~eV above midgap) is considered with $D_{\rm it}=2.5\times 10^{11}$cm$^{-2}$eV$^{-1}$. The channel is undoped with electron mobility $\mu=85$cm$^2$/Vs, saturation velocity $v_{\rm sat}=2.8\cdot 10^6$cm/s, electron effective mass  $m^*=0.61 m_0$ and bandgap width $E_{g}=1.8$eV. The experimental data and the simulation results are depicted in \figref{fig:ajusteidsvfg}, showing a very good agreement with the  bias point employed in the RF experimental characterization, $V_{\rm GS,e}=-3$~V and $V_{\rm DS,e} = 2$~V, marked by a square. Remarkably, at this bias point the resistance associated with the access regions (as computed from the self-consistent simulations),  have comparable values to $R_{\rm c}$, reaching $R_{\rm s, acc}=3.3$~k$\Omega\cdot\mu$m and $R_{\rm d, acc}=3.1$~k$\Omega\cdot\mu$m at the source and drain ends, respectively and being rather bias dependent. This point highlights the strong impact that these gate-underlapped areas can produce on the RF performance of the device \cite{Toral-Lopez2019a}.

Using these results, we can assess the proposed approach by confronting the RF figures of merit resulting from it with those experimentally characterized and reported in \cite{Krasnozhon2014}.
To this purpose, the values of the small-signal parameters of the device are first extracted following the procedure previously described and fed into the small-signal model (\tabref{tab:smallsigparam}). 

\begin{table}[th]
	\centering
	\begin{tabular}{cc|cc}
		\hline
		\textbf{Parameter} & \textbf{Value} & \textbf{Parameter} & \textbf{Value} \\ 
		\hline \hline 
		$g_{\rm m}$ & 0.0828 mS/$\mu$m  & $g_{\rm ds}$ & 0.5047 mS/$\mu$m \\ 
		
		$C_{\rm gd}$ & 0.280 fF/$\mu$m & $C_{\rm dg}$ & 0.442 fF/$\mu$m \\ 
		
		$C_{\rm sd}$ & -0.071 fF/$\mu$m & $C_{\rm gs}$ & 0.604 fF/$\mu$m \\ 
		
		\hline \hline 
	\end{tabular} 
	\caption{Small-signal parameters extracted from the self-consistent numerical simulator.}
	\label{tab:smallsigparam}
\end{table}

\begin{figure}[th]
	\centering
	\includegraphics[width=\linewidth]{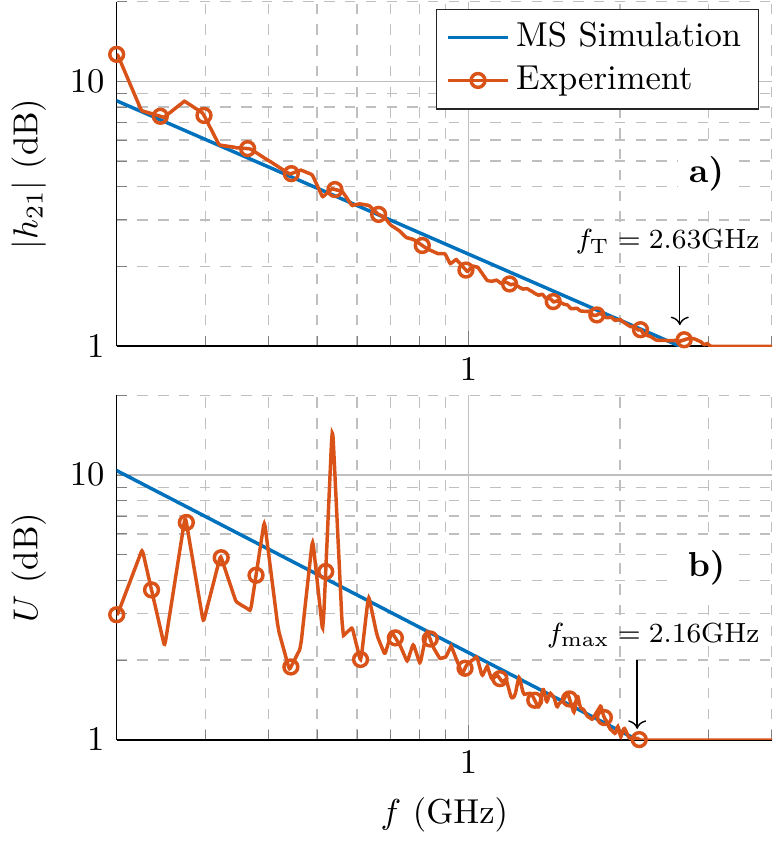}
	\caption{a) Current gain $h_{21}$ and b) Mason's unilateral gain $U$ calculated using the multi-scale (MS) approach (blue lines) and compared against the experimental values extracted from \cite{Krasnozhon2014} (red lines with symbols). The arrows indicate the values of $f_{\rm T}$ and $f_{\rm max}$.}
	\label{fig:freqh21u}
\end{figure}

The current gain ($h_{21}$) and the Mason's unilateral gain ($U$) obtained from the multi-scale scheme, i.e. combining the small-signal-model and numerical-solver analysis, are depicted in \figref{fig:freqh21u} along with the experimental data \cite{Krasnozhon2014}, demonstrating an excellent correspondence and holding the soundness of the theoretical procedure. $f_{\rm T}$ and $f_{\rm max}$, obtained from the numerical calculations, marked within the figures, are also shown in very close agreement with the reported measurements: $f_{\rm T} = 2.63$GHz and $f_{\rm max} = 2.16$GHz.

Once  the multi-scale approach is validated, we exemplify its utility by proceeding with a projective analysis of the potential of MoS$_2$ FETs for RF applications; in particular extracting $f_{\rm T}$ and $f_{\rm max}$ for devices with channel lengths ranging from 10~$\mu$m down to $50$~nm. In order to focus on the material intrinsic capabilities and to reduce the role of extrinsic resistive elements in the predicted behavior, the 2D-FET is improved by reducing the access regions length to 5nm and setting contact resistances to $100\Omega\cdot\mu$m, which is the desirable value for 2DMs to compete on the RF arena \cite{Jena2014} {and not so far from the minimum value achieved experimentally, 240 $\Omega\cdot\mu$m \cite{Kappera2014}, and well above the theoretical limit of 30 $\Omega\cdot\mu$m \cite{Allain_2015}.}

\begin{figure}[th]
	\centering
	\includegraphics[width=\linewidth]{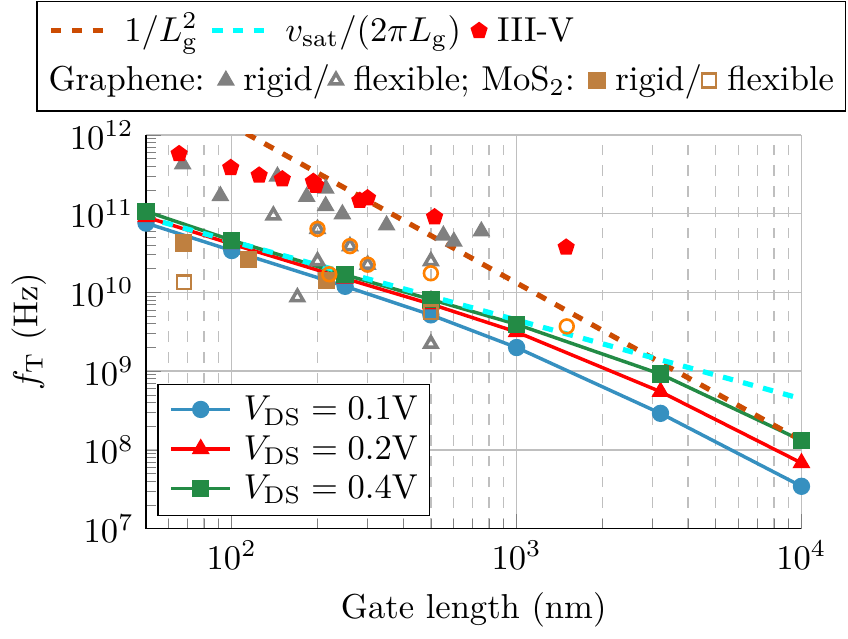}
	\caption{Cut-off frequency $f_{\rm T}$ as a function of the gate length for different drain biases (solid lines with markers). The $1/L_{\rm g}^{2}$ trend and the physical limit (that also serves as a guideline for the $1/L_{\rm g}$ scaling) are also depicted (dashed lines). Solid and hollow markers correspond to experimental works  \cite{Schwierz2013,Chang2015,Cheng2014,Park2016,Lee2013,Sire2012,Petrone2015,Lee2012,Park2017,Wei2016,Yeh2014}.}
	\label{fig:ftlchn}
\end{figure}

\figref{fig:ftlchn} shows $f_{\rm T}$ as a function of the gate length, $L_{\rm g}$. Each marker of the solid curves corresponds to the result achieved from the application of the multi-scale procedure to a device with a particular $L_{\rm g}$ under a given $V_{\rm DS}$ and considering the $V_{\rm GS}$ bias that causes the maximum transconductance. Along with these results, we show the physical limit represented by $v_{\rm sat}/2\pi L_{\rm g}$, i.e. the inverse of the minimum transit time of charge carriers in the channel, that also serves as a guide for the eyes of the $1/L_{\rm g}$ scaling; and the $1/L_{\rm g}^2$ trend that conventionally commands the performance in longer channels. We also show, for the sake of comparison, the results for different experimental devices based on graphene, MoS$_2$ and III-V compounds taken from the literature (solid and hollow symbols) \cite{Schwierz2013,Chang2015,Cheng2014,Park2016,Lee2013,Sire2012,Petrone2015,Lee2012,Park2017,Wei2016,Yeh2014}. 

According to \figref{fig:ftlchn}, a change in the scaling trend of MoS$_2$ $f_{\rm T}$, from $1/L_{\rm g}^2$  for long channels to $1/L_{\rm g}$ for short channels can be expected. This can be explained due to the $f_{\rm T}$ rough dependence on $\sim g_{\rm m}/2\pi(C_{\rm gs}+C_{\rm gd})$. For long channels, $g_{\rm m}$ is proportional to $1/L_{\rm g}$, and $(C_{\rm gs}+C_{\rm gd})$ to $L_{\rm g}$, resulting in a trend $\sim1/ L_{\rm g}^2$; while for short channels $g_{\rm m}$ saturates and $f_{\rm T}$ scales only with the capacitive response, to $1/L_{\rm g}$. The transition from $1/L_{\rm g}^2$ to $1/L_{\rm g}$ is observed in MoS$_2$ still for long channels, i.e. $L_{\rm g}\approx 1\mu$m due to $g_{\rm m}$ early saturation \cite{Schwierz2013}. The $v_{\rm sat}$ impact is also revealed in the dependence of $f_{\rm T}$ on $V_{\rm DS}$: when the channel length is scaled, the performance is barely improved with $V_{\rm DS}$, as it approaches the physical limit, i.e., the carrier drift velocity cannot increase any further with the electric field. The values predicted are in good agreement with experimental state-of-the-art RF measurements of MoS$_2$ devices, demonstrating the prediction capability of the presented tool, that can be exploited to study any candidate in the 2DM realm. The experimental results shown for graphene and III-V compounds contextualize the progresses already achieved by MoS$_2$ electronics and the limitations associated to its lower saturation velocity. 

\begin{figure}[th]
	\centering
	\includegraphics[width=\linewidth]{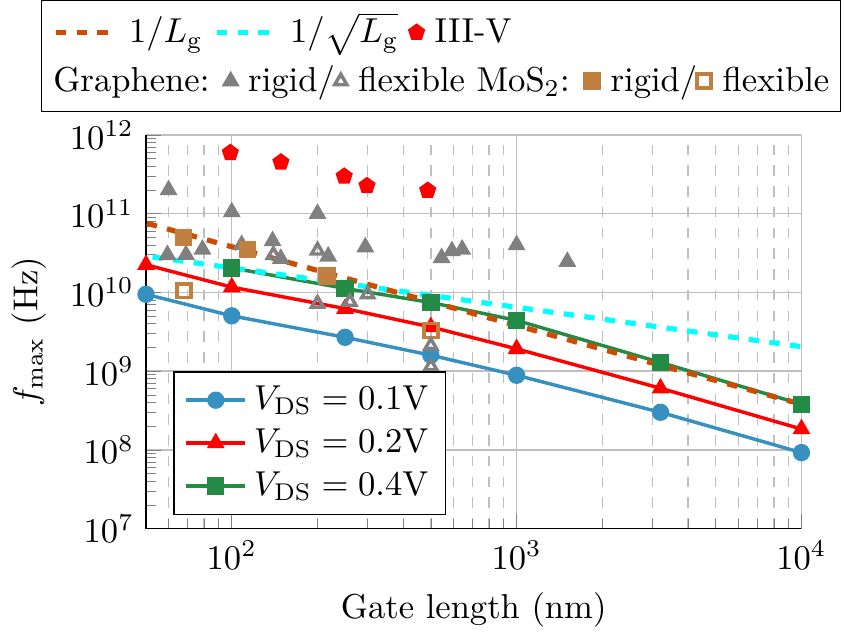}	
	\caption{Maximum oscillation frequency $f_{\rm max}$ as a function of the gate length for different drain biases (solid lines with markers). The $1/L_{\rm g}$ and $1/\sqrt{L_{\rm g}}$ trends are indicated by  dashed lines. Solid and hollow markers correspond to experimental works  \cite{Schwierz2013,Chang2015,Cheng2014,Park2016,Lee2013,Sire2012,Petrone2015,Lee2012,Park2017,Wei2016,Yeh2014}.}
	\label{fig:fmaxlchn}
\end{figure}

Concerning $f_{\rm max}$, \figref{fig:fmaxlchn} plots it as a function of $L_{\rm g}$ for different $V_{\rm DS}$ biases, together with the $1/L_{\rm g}$ and $1/\sqrt{L_{\rm g}}$ lines used as a guide for the eyes, and the experimental results from different technologies. For long channel devices, $f_{\rm max}$  varies with $1/ L_{\rm g}$  as $f_{\rm max}$ depends roughly on $\sim \sqrt{f_{\rm T}/(8\pi R_{\rm g}C_{\rm gd})}$ considering that the gate resistance $R_{\rm g}$ si proportional to $1/ L_{\rm g}$ \cite{Sze2006}. For short channel devices, $L_{\rm g} < 500$ nm, the trend changes from $1/ L_{\rm g}$ to $1/ \sqrt{L_{\rm g}}$, showing a small over-increment at the limit of the diffusive regime, as it has also  been predicted for GFETs in \cite{Feijoo2017}. Although the experimental values correspond to diverse technologies and measured under different operating bias points, the measurements show good agreement with our predictions. More importantly, MoS$_2$ FETs demonstrate to be competitive or even to overcome graphene devices performance for the smaller channel lengths, thanks to their larger output conductance, postulating MoS$_2$ as a worthy 2D alternative for RF power circuits, being nevertheless still far from consolidated III-V architectures.

\section{Conclusion}

We reported a multi-scale approach that combines small-signal and numerical simulations in order to describe in detail the behavior of 2DM-based devices for RF applications. Both levels of abstraction are precisely combined, so the main features included in the Poisson-Drift-diffusion system are extended to circuit level simulations. The proposed scheme was validated against experimental data, showing an excellent agreement with dc and RF measurements. The multi-scale simulation tool was then employed to project the impact of channel length scaling on the RF performance of a MoS$_2$ based FET with improved extrinsic behavior by considering a reduction in the contact resistance and shorter gate under-lapped access regions, so to understand and assess the intrinsic limits of the 2DMs. The resulting $f_{\rm T}$ vs. $L_{\rm g}$ curves show a change in their trend from $1/L_{\rm g}^2$ to $1/L_{\rm g}$ and a saturated behavior with respect to $V_{\rm DS}$ for low $L_{\rm g}$, due to the constraining saturation velocity. When comparing the results  with those achieved in experimental devices, the reduction of the access regions and contact resistances implies an increase of almost 10-fold in $f_{\rm T}$ with respect to the experimental sample \cite{Sanne2015, Krasnozhon2014}. On the other hand, the $f_{\rm max}$ vs. $L_{\rm g}$ curves present a scaling-trend varying from $1/L_{\rm g}$ to $1/\sqrt{L_{\rm g}}$, that is impacted by the up-scaling of the gate resistance.  An improvement of the RF performance of MoS$_2$-based devices could be achieved by enhancing the crystal transport properties, e.g., higher mobility and saturation velocity. Hence, the multi-level approach presented here constitutes a versatile platform for evaluating the impact of scaling on the RF performance of any 2D material based FET. {In addition to channel scaling, the capabilities of this approach can be extended to asses the impact of surface defects and even mechanical strain in devices fabricated on flexible substrates}. Furthermore, it enables the design and assessment of sophisticated RF circuits based on such devices. 

\section*{Conflicts of interest}
Authors do not have conflicts of interest to declare.

\section*{Acknowledgements}
A. Toral-Lopez acknowledges the FPU program (FPU16/04043). E.G. Marin acknowledges Juan de la Cierva Incorporación IJCI-2017-32297 (MINECO/AEI). This work has received funding from the European Unions Horizon 2020 research and innovation programme under grant agreements No GrapheneCore2 785219 and No GrapheneCore3 881603, and from Ministerio de Ciencia, Innovación y Universidades under grant agreement RTI2018-097876-B-C21(MCIU/AEI/FEDER, UE), TEC2017-89955-P (MINECO/AEI/FEDER, UE), and TEC2015-67462-C2-1-R (MINECO). This article has been partially funded by the European Regional Development Funds (ERDF) allocated to the Programa Operatiu FEDER de Catalunya 2014-2020, with support of the Secretaria d'Universitats i Recerca of the Departament d'Empresa i Coneixement of the Generalitat de Catalunya for emerging technology clusters to carry out valorization and transfer of research results. GraphCAT project 001-P-001702.




\renewcommand\refname{References}

\bibliographystyle{rsc} 
\bibliography{MScale_MOS-RF_2DMat}

\providecommand*{\mcitethebibliography}{\thebibliography}
\csname @ifundefined\endcsname{endmcitethebibliography}
{\let\endmcitethebibliography\endthebibliography}{}
\begin{mcitethebibliography}{43}
\providecommand*{\natexlab}[1]{#1}
\providecommand*{\mciteSetBstSublistMode}[1]{}
\providecommand*{\mciteSetBstMaxWidthForm}[2]{}
\providecommand*{\mciteBstWouldAddEndPuncttrue}
  {\def\EndOfBibitem{\unskip.}}
\providecommand*{\mciteBstWouldAddEndPunctfalse}
  {\let\EndOfBibitem\relax}
\providecommand*{\mciteSetBstMidEndSepPunct}[3]{}
\providecommand*{\mciteSetBstSublistLabelBeginEnd}[3]{}
\providecommand*{\EndOfBibitem}{}
\mciteSetBstSublistMode{f}
\mciteSetBstMaxWidthForm{subitem}
{(\emph{\alph{mcitesubitemcount}})}
\mciteSetBstSublistLabelBeginEnd{\mcitemaxwidthsubitemform\space}
{\relax}{\relax}

\bibitem[Cheng \emph{et~al.}(2012)Cheng, Bai, Liao, Zhou, Chen, Liu, Lin,
  Jiang, Huang, and Duan]{Cheng2012}
R.~Cheng, J.~Bai, L.~Liao, H.~Zhou, Y.~Chen, L.~Liu, Y.-C. Lin, S.~Jiang,
  Y.~Huang and X.~Duan, \emph{Proceedings of the National Academy of Sciences},
  2012, \textbf{109}, 11588--11592\relax
\mciteBstWouldAddEndPuncttrue
\mciteSetBstMidEndSepPunct{\mcitedefaultmidpunct}
{\mcitedefaultendpunct}{\mcitedefaultseppunct}\relax
\EndOfBibitem
\bibitem[Liao \emph{et~al.}(2010)Liao, Lin, Bao, Cheng, Bai, Liu, Qu, Wang,
  Huang, and Duan]{Liao2010}
L.~Liao, Y.-C. Lin, M.~Bao, R.~Cheng, J.~Bai, Y.~Liu, Y.~Qu, K.~L. Wang,
  Y.~Huang and X.~Duan, \emph{Nature}, 2010, \textbf{467}, 305--308\relax
\mciteBstWouldAddEndPuncttrue
\mciteSetBstMidEndSepPunct{\mcitedefaultmidpunct}
{\mcitedefaultendpunct}{\mcitedefaultseppunct}\relax
\EndOfBibitem
\bibitem[Wu \emph{et~al.}(2016)Wu, Zou, Sun, Cao, Wang, Huo, Zhou, Yang, Yu,
  Kong, Yu, Liao, and Chen]{Wu2016}
Y.~Wu, X.~Zou, M.~Sun, Z.~Cao, X.~Wang, S.~Huo, J.~Zhou, Y.~Yang, X.~Yu,
  Y.~Kong, G.~Yu, L.~Liao and T.~Chen, \emph{{ACS} Applied Materials {\&}
  Interfaces}, 2016, \textbf{8}, 25645--25649\relax
\mciteBstWouldAddEndPuncttrue
\mciteSetBstMidEndSepPunct{\mcitedefaultmidpunct}
{\mcitedefaultendpunct}{\mcitedefaultseppunct}\relax
\EndOfBibitem
\bibitem[Feng \emph{et~al.}(2014)Feng, Yu, Li, Liu, He, Song, Wang, and
  Cai]{Feng2014}
Z.~Feng, C.~Yu, J.~Li, Q.~Liu, Z.~He, X.~Song, J.~Wang and S.~Cai,
  \emph{Carbon}, 2014, \textbf{75}, 249--254\relax
\mciteBstWouldAddEndPuncttrue
\mciteSetBstMidEndSepPunct{\mcitedefaultmidpunct}
{\mcitedefaultendpunct}{\mcitedefaultseppunct}\relax
\EndOfBibitem
\bibitem[Guo \emph{et~al.}(2013)Guo, Dong, Chakraborty, Lourenco, Palmer, Hu,
  Ruan, Hankinson, Kunc, Cressler, Berger, and de~Heer]{Guo2013}
Z.~Guo, R.~Dong, P.~S. Chakraborty, N.~Lourenco, J.~Palmer, Y.~Hu, M.~Ruan,
  J.~Hankinson, J.~Kunc, J.~D. Cressler, C.~Berger and W.~A. de~Heer,
  \emph{Nano Letters}, 2013, \textbf{13}, 942--947\relax
\mciteBstWouldAddEndPuncttrue
\mciteSetBstMidEndSepPunct{\mcitedefaultmidpunct}
{\mcitedefaultendpunct}{\mcitedefaultseppunct}\relax
\EndOfBibitem
\bibitem[Schwierz(2013)]{Schwierz2013}
F.~Schwierz, \emph{Proceedings of the {IEEE}}, 2013, \textbf{101},
  1567--1584\relax
\mciteBstWouldAddEndPuncttrue
\mciteSetBstMidEndSepPunct{\mcitedefaultmidpunct}
{\mcitedefaultendpunct}{\mcitedefaultseppunct}\relax
\EndOfBibitem
\bibitem[Krasnozhon \emph{et~al.}(2014)Krasnozhon, Lembke, Nyffeler, Leblebici,
  and Kis]{Krasnozhon2014}
D.~Krasnozhon, D.~Lembke, C.~Nyffeler, Y.~Leblebici and A.~Kis, \emph{Nano
  Letters}, 2014, \textbf{14}, 5905--5911\relax
\mciteBstWouldAddEndPuncttrue
\mciteSetBstMidEndSepPunct{\mcitedefaultmidpunct}
{\mcitedefaultendpunct}{\mcitedefaultseppunct}\relax
\EndOfBibitem
\bibitem[Gao \emph{et~al.}(2018)Gao, Zhang, Xu, Song, Li, and Wu]{Gao2018}
Q.~Gao, Z.~Zhang, X.~Xu, J.~Song, X.~Li and Y.~Wu, \emph{Nature
  Communications}, 2018, \textbf{9}, year\relax
\mciteBstWouldAddEndPuncttrue
\mciteSetBstMidEndSepPunct{\mcitedefaultmidpunct}
{\mcitedefaultendpunct}{\mcitedefaultseppunct}\relax
\EndOfBibitem
\bibitem[Chang \emph{et~al.}(2015)Chang, Yogeesh, Ghosh, Rai, Sanne, Yang, Lu,
  Banerjee, and Akinwande]{Chang2015}
H.-Y. Chang, M.~N. Yogeesh, R.~Ghosh, A.~Rai, A.~Sanne, S.~Yang, N.~Lu, S.~K.
  Banerjee and D.~Akinwande, \emph{Advanced Materials}, 2015, \textbf{28},
  1818--1823\relax
\mciteBstWouldAddEndPuncttrue
\mciteSetBstMidEndSepPunct{\mcitedefaultmidpunct}
{\mcitedefaultendpunct}{\mcitedefaultseppunct}\relax
\EndOfBibitem
\bibitem[Sanne \emph{et~al.}(2015)Sanne, Ghosh, Rai, Yogeesh, Shin, Sharma,
  Jarvis, Mathew, Rao, Akinwande, and Banerjee]{Sanne2015}
A.~Sanne, R.~Ghosh, A.~Rai, M.~N. Yogeesh, S.~H. Shin, A.~Sharma, K.~Jarvis,
  L.~Mathew, R.~Rao, D.~Akinwande and S.~Banerjee, \emph{Nano Letters}, 2015,
  \textbf{15}, 5039--5045\relax
\mciteBstWouldAddEndPuncttrue
\mciteSetBstMidEndSepPunct{\mcitedefaultmidpunct}
{\mcitedefaultendpunct}{\mcitedefaultseppunct}\relax
\EndOfBibitem
\bibitem[Sanne \emph{et~al.}(2017)Sanne, Park, Ghosh, Yogeesh, Liu, Mathew,
  Rao, Akinwande, and Banerjee]{Sanne2017}
A.~Sanne, S.~Park, R.~Ghosh, M.~N. Yogeesh, C.~Liu, L.~Mathew, R.~Rao,
  D.~Akinwande and S.~K. Banerjee, \emph{npj {2D} Materials and Applications},
  2017, \textbf{1}, year\relax
\mciteBstWouldAddEndPuncttrue
\mciteSetBstMidEndSepPunct{\mcitedefaultmidpunct}
{\mcitedefaultendpunct}{\mcitedefaultseppunct}\relax
\EndOfBibitem
\bibitem[Cheng \emph{et~al.}(2014)Cheng, Jiang, Chen, Liu, Weiss, Cheng, Wu,
  Huang, and Duan]{Cheng2014}
R.~Cheng, S.~Jiang, Y.~Chen, Y.~Liu, N.~Weiss, H.~C. Cheng, H.~Wu, Y.~Huang and
  X.~Duan, \emph{Nature Communications}, 2014, \textbf{5}, 1--9\relax
\mciteBstWouldAddEndPuncttrue
\mciteSetBstMidEndSepPunct{\mcitedefaultmidpunct}
{\mcitedefaultendpunct}{\mcitedefaultseppunct}\relax
\EndOfBibitem
\bibitem[Belete \emph{et~al.}(2018)Belete, Kataria, Koch, Kruth, Engelhard,
  Mayer, Engström, and Lemme]{Belete2018}
M.~Belete, S.~Kataria, U.~Koch, M.~Kruth, C.~Engelhard, J.~Mayer, O.~Engström
  and M.~C. Lemme, \emph{{ACS} Applied Nano Materials}, 2018, \textbf{1},
  6197--6204\relax
\mciteBstWouldAddEndPuncttrue
\mciteSetBstMidEndSepPunct{\mcitedefaultmidpunct}
{\mcitedefaultendpunct}{\mcitedefaultseppunct}\relax
\EndOfBibitem
\bibitem[Bartolomeo \emph{et~al.}(2019)Bartolomeo, Pelella, Liu, Miao,
  Passacantando, Giubileo, Grillo, Iemmo, Urban, and Liang]{Bartolomeo2019}
A.~D. Bartolomeo, A.~Pelella, X.~Liu, F.~Miao, M.~Passacantando, F.~Giubileo,
  A.~Grillo, L.~Iemmo, F.~Urban and S.-J. Liang, \emph{Advanced Functional
  Materials}, 2019, \textbf{29}, 1902483\relax
\mciteBstWouldAddEndPuncttrue
\mciteSetBstMidEndSepPunct{\mcitedefaultmidpunct}
{\mcitedefaultendpunct}{\mcitedefaultseppunct}\relax
\EndOfBibitem
\bibitem[Urban \emph{et~al.}(2020)Urban, Gity, Hurley, McEvoy, and
  Bartolomeo]{Urban2020}
F.~Urban, F.~Gity, P.~K. Hurley, N.~McEvoy and A.~D. Bartolomeo, \emph{Applied
  Physics Letters}, 2020, \textbf{117}, 193102\relax
\mciteBstWouldAddEndPuncttrue
\mciteSetBstMidEndSepPunct{\mcitedefaultmidpunct}
{\mcitedefaultendpunct}{\mcitedefaultseppunct}\relax
\EndOfBibitem
\bibitem[Schwierz \emph{et~al.}(2015)Schwierz, Pezoldt, and
  Granzner]{Schwierz2015}
F.~Schwierz, J.~Pezoldt and R.~Granzner, \emph{Nanoscale}, 2015, \textbf{7},
  8261--8283\relax
\mciteBstWouldAddEndPuncttrue
\mciteSetBstMidEndSepPunct{\mcitedefaultmidpunct}
{\mcitedefaultendpunct}{\mcitedefaultseppunct}\relax
\EndOfBibitem
\bibitem[Marin \emph{et~al.}(2018)Marin, Perucchini, Marian, Iannaccone, and
  Fiori]{Marin2018b}
E.~G. Marin, M.~Perucchini, D.~Marian, G.~Iannaccone and G.~Fiori, \emph{IEEE
  Transactions on Electron Devices}, 2018\relax
\mciteBstWouldAddEndPuncttrue
\mciteSetBstMidEndSepPunct{\mcitedefaultmidpunct}
{\mcitedefaultendpunct}{\mcitedefaultseppunct}\relax
\EndOfBibitem
\bibitem[Pasadas \emph{et~al.}(2019)Pasadas, Marin, Toral-Lopez, Ruiz, Godoy,
  Park, Akinwande, and Jim{\'{e}}nez]{Pasadas2019}
F.~Pasadas, E.~G. Marin, A.~Toral-Lopez, F.~G. Ruiz, A.~Godoy, S.~Park,
  D.~Akinwande and D.~Jim{\'{e}}nez, \emph{npj {2D} Materials and
  Applications}, 2019\relax
\mciteBstWouldAddEndPuncttrue
\mciteSetBstMidEndSepPunct{\mcitedefaultmidpunct}
{\mcitedefaultendpunct}{\mcitedefaultseppunct}\relax
\EndOfBibitem
\bibitem[Bartolomeo \emph{et~al.}(2017)Bartolomeo, Genovese, Giubileo, Iemmo,
  Luongo, Foller, and Schleberger]{Bartolomeo2017}
A.~D. Bartolomeo, L.~Genovese, F.~Giubileo, L.~Iemmo, G.~Luongo, T.~Foller and
  M.~Schleberger, \emph{2D Materials}, 2017, \textbf{5}, 015014\relax
\mciteBstWouldAddEndPuncttrue
\mciteSetBstMidEndSepPunct{\mcitedefaultmidpunct}
{\mcitedefaultendpunct}{\mcitedefaultseppunct}\relax
\EndOfBibitem
\bibitem[Feijoo \emph{et~al.}(2019)Feijoo, Pasadas, Iglesias, Hamham, Rengel,
  and Jimenez]{Feijoo2019}
P.~C. Feijoo, F.~Pasadas, J.~M. Iglesias, E.~M. Hamham, R.~Rengel and
  D.~Jimenez, \emph{{IEEE} Transactions on Electron Devices}, 2019,  1--7\relax
\mciteBstWouldAddEndPuncttrue
\mciteSetBstMidEndSepPunct{\mcitedefaultmidpunct}
{\mcitedefaultendpunct}{\mcitedefaultseppunct}\relax
\EndOfBibitem
\bibitem[Feijoo \emph{et~al.}(2016)Feijoo, Jim{\'{e}}nez, and
  Cartoix{\`{a}}]{Feijoo_2016}
P.~C. Feijoo, D.~Jim{\'{e}}nez and X.~Cartoix{\`{a}}, \emph{{2D} Materials},
  2016, \textbf{3}, 025036\relax
\mciteBstWouldAddEndPuncttrue
\mciteSetBstMidEndSepPunct{\mcitedefaultmidpunct}
{\mcitedefaultendpunct}{\mcitedefaultseppunct}\relax
\EndOfBibitem
\bibitem[Pasadas and Jimenez(2016)]{Pasadas2016}
F.~Pasadas and D.~Jimenez, \emph{{IEEE} Transactions on Electron Devices},
  2016, \textbf{63}, 2936--2941\relax
\mciteBstWouldAddEndPuncttrue
\mciteSetBstMidEndSepPunct{\mcitedefaultmidpunct}
{\mcitedefaultendpunct}{\mcitedefaultseppunct}\relax
\EndOfBibitem
\bibitem[Ward and Dutton(1978)]{Ward1978}
D.~Ward and R.~Dutton, \emph{IEEE Journal of Solid-State Circuits}, 1978,
  \textbf{13}, 703--708\relax
\mciteBstWouldAddEndPuncttrue
\mciteSetBstMidEndSepPunct{\mcitedefaultmidpunct}
{\mcitedefaultendpunct}{\mcitedefaultseppunct}\relax
\EndOfBibitem
\bibitem[Pasadas \emph{et~al.}(2017)Pasadas, Wei, Pallecchi, Happy, and
  Jimenez]{Pasadas2017}
F.~Pasadas, W.~Wei, E.~Pallecchi, H.~Happy and D.~Jimenez, \emph{{IEEE}
  Transactions on Electron Devices}, 2017, \textbf{64}, 4715--4723\relax
\mciteBstWouldAddEndPuncttrue
\mciteSetBstMidEndSepPunct{\mcitedefaultmidpunct}
{\mcitedefaultendpunct}{\mcitedefaultseppunct}\relax
\EndOfBibitem
\bibitem[Jena \emph{et~al.}(2014)Jena, Banerjee, and Xing]{Jena2014}
D.~Jena, K.~Banerjee and G.~H. Xing, \emph{Nature Materials}, 2014,
  \textbf{13}, 1076--1078\relax
\mciteBstWouldAddEndPuncttrue
\mciteSetBstMidEndSepPunct{\mcitedefaultmidpunct}
{\mcitedefaultendpunct}{\mcitedefaultseppunct}\relax
\EndOfBibitem
\bibitem[Meyer()]{Meyer}
J.~E. Meyer, \emph{MOS models and circuit simulation}, RCA Rev, pp.
  42--63\relax
\mciteBstWouldAddEndPuncttrue
\mciteSetBstMidEndSepPunct{\mcitedefaultmidpunct}
{\mcitedefaultendpunct}{\mcitedefaultseppunct}\relax
\EndOfBibitem
\bibitem[Toral-Lopez \emph{et~al.}(2019)Toral-Lopez, Pasadas, Marin,
  Medina-Rull, Ruiz, Jimenez, and Godoy]{ToralLopez2019}
A.~Toral-Lopez, F.~Pasadas, E.~G. Marin, A.~Medina-Rull, F.~J.~G. Ruiz,
  D.~Jimenez and A.~Godoy, 2019 International Conference on Simulation of
  Semiconductor Processes and Devices (SISPAD), 2019, pp. 1--4\relax
\mciteBstWouldAddEndPuncttrue
\mciteSetBstMidEndSepPunct{\mcitedefaultmidpunct}
{\mcitedefaultendpunct}{\mcitedefaultseppunct}\relax
\EndOfBibitem
\bibitem[Dagan \emph{et~al.}(2020)Dagan, Vaknin, and Rosenwaks]{Dagan2020}
R.~Dagan, Y.~Vaknin and Y.~Rosenwaks, \emph{Nanoscale}, 2020, \textbf{12},
  8883--8889\relax
\mciteBstWouldAddEndPuncttrue
\mciteSetBstMidEndSepPunct{\mcitedefaultmidpunct}
{\mcitedefaultendpunct}{\mcitedefaultseppunct}\relax
\EndOfBibitem
\bibitem[Kim \emph{et~al.}(2017)Kim, Moon, Lee, Choi, Ahmed, Nam, Cho, Shin,
  Park, and Yoo]{Kim2017}
C.~Kim, I.~Moon, D.~Lee, M.~S. Choi, F.~Ahmed, S.~Nam, Y.~Cho, H.-J. Shin,
  S.~Park and W.~J. Yoo, \emph{{ACS} Nano}, 2017, \textbf{11}, 1588--1596\relax
\mciteBstWouldAddEndPuncttrue
\mciteSetBstMidEndSepPunct{\mcitedefaultmidpunct}
{\mcitedefaultendpunct}{\mcitedefaultseppunct}\relax
\EndOfBibitem
\bibitem[Takenaka \emph{et~al.}(2016)Takenaka, Ozawa, Han, and
  Takagi]{Takenaka2016}
M.~Takenaka, Y.~Ozawa, J.~Han and S.~Takagi, 2016 {IEEE} International Electron
  Devices Meeting ({IEDM}), 2016\relax
\mciteBstWouldAddEndPuncttrue
\mciteSetBstMidEndSepPunct{\mcitedefaultmidpunct}
{\mcitedefaultendpunct}{\mcitedefaultseppunct}\relax
\EndOfBibitem
\bibitem[Toral-Lopez \emph{et~al.}(2019)Toral-Lopez, Marin, Medina, Ruiz,
  Rodriguez, and Godoy]{Toral-Lopez2019a}
A.~Toral-Lopez, E.~G. Marin, A.~Medina, F.~G. Ruiz, N.~Rodriguez and A.~Godoy,
  \emph{Nanomaterials}, 2019, \textbf{9}, 1027\relax
\mciteBstWouldAddEndPuncttrue
\mciteSetBstMidEndSepPunct{\mcitedefaultmidpunct}
{\mcitedefaultendpunct}{\mcitedefaultseppunct}\relax
\EndOfBibitem
\bibitem[Kappera \emph{et~al.}(2014)Kappera, Voiry, Yalcin, Branch, Gupta,
  Mohite, and Chhowalla]{Kappera2014}
R.~Kappera, D.~Voiry, S.~E. Yalcin, B.~Branch, G.~Gupta, A.~D. Mohite and
  M.~Chhowalla, \emph{Nature Materials}, 2014, \textbf{13}, 1128--1134\relax
\mciteBstWouldAddEndPuncttrue
\mciteSetBstMidEndSepPunct{\mcitedefaultmidpunct}
{\mcitedefaultendpunct}{\mcitedefaultseppunct}\relax
\EndOfBibitem
\bibitem[Allain \emph{et~al.}(2015)Allain, Kang, Banerjee, and
  Kis]{Allain_2015}
A.~Allain, J.~Kang, K.~Banerjee and A.~Kis, \emph{Nature Materials}, 2015,
  \textbf{14}, 1195--1205\relax
\mciteBstWouldAddEndPuncttrue
\mciteSetBstMidEndSepPunct{\mcitedefaultmidpunct}
{\mcitedefaultendpunct}{\mcitedefaultseppunct}\relax
\EndOfBibitem
\bibitem[Park \emph{et~al.}(2016)Park, Shin, Yogeesh, Lee, Rahimi, and
  Akinwande]{Park2016}
S.~Park, S.~H. Shin, M.~N. Yogeesh, A.~L. Lee, S.~Rahimi and D.~Akinwande,
  \emph{IEEE Electron Device Letters}, 2016, \textbf{37}, 512--515\relax
\mciteBstWouldAddEndPuncttrue
\mciteSetBstMidEndSepPunct{\mcitedefaultmidpunct}
{\mcitedefaultendpunct}{\mcitedefaultseppunct}\relax
\EndOfBibitem
\bibitem[Lee \emph{et~al.}(2013)Lee, Ha, Li, Parrish, Holt, Dodabalapur, Ruoff,
  and Akinwande]{Lee2013}
J.~Lee, T.-J. Ha, H.~Li, K.~N. Parrish, M.~Holt, A.~Dodabalapur, R.~S. Ruoff
  and D.~Akinwande, \emph{ACS Nano}, 2013, \textbf{7}, 7744--7750\relax
\mciteBstWouldAddEndPuncttrue
\mciteSetBstMidEndSepPunct{\mcitedefaultmidpunct}
{\mcitedefaultendpunct}{\mcitedefaultseppunct}\relax
\EndOfBibitem
\bibitem[Sire \emph{et~al.}(2012)Sire, Ardiaca, Lepilliet, Seo, Hersam,
  Dambrine, Happy, and Derycke]{Sire2012}
C.~Sire, F.~Ardiaca, S.~Lepilliet, J.-W.~T. Seo, M.~C. Hersam, G.~Dambrine,
  H.~Happy and V.~Derycke, \emph{Nano Letters}, 2012, \textbf{12},
  1184--1188\relax
\mciteBstWouldAddEndPuncttrue
\mciteSetBstMidEndSepPunct{\mcitedefaultmidpunct}
{\mcitedefaultendpunct}{\mcitedefaultseppunct}\relax
\EndOfBibitem
\bibitem[Petrone \emph{et~al.}(2015)Petrone, Meric, Chari, Shepard, and
  Hone]{Petrone2015}
N.~Petrone, I.~Meric, T.~Chari, K.~L. Shepard and J.~Hone, \emph{IEEE Journal
  of the Electron Devices Society}, 2015, \textbf{3}, 44--48\relax
\mciteBstWouldAddEndPuncttrue
\mciteSetBstMidEndSepPunct{\mcitedefaultmidpunct}
{\mcitedefaultendpunct}{\mcitedefaultseppunct}\relax
\EndOfBibitem
\bibitem[Lee \emph{et~al.}(2012)Lee, Parrish, Chowdhury, Ha, Hao, Tao,
  Dodabalapur, Ruoff, and Akinwande]{Lee2012}
J.~Lee, K.~N. Parrish, S.~F. Chowdhury, T.-J. Ha, Y.~Hao, L.~Tao,
  A.~Dodabalapur, R.~S. Ruoff and D.~Akinwande, 2012 International Electron
  Devices Meeting, 2012, pp. 14.6.1--14.6.4\relax
\mciteBstWouldAddEndPuncttrue
\mciteSetBstMidEndSepPunct{\mcitedefaultmidpunct}
{\mcitedefaultendpunct}{\mcitedefaultseppunct}\relax
\EndOfBibitem
\bibitem[Park and Akinwande(2017)]{Park2017}
S.~Park and D.~Akinwande, 2017 IEEE International Electron Devices Meeting
  (IEDM), 2017, pp. 5.2.1--5.2.4\relax
\mciteBstWouldAddEndPuncttrue
\mciteSetBstMidEndSepPunct{\mcitedefaultmidpunct}
{\mcitedefaultendpunct}{\mcitedefaultseppunct}\relax
\EndOfBibitem
\bibitem[Wei \emph{et~al.}(2016)Wei, Pallecchi, Belhaj, Centeno, Amaia,
  Vignaud, and Happy]{Wei2016}
W.~Wei, E.~Pallecchi, M.~Belhaj, A.~Centeno, Z.~Amaia, D.~Vignaud and H.~Happy,
  2016 11th European Microwave Integrated Circuits Conference (EuMIC), 2016,
  pp. 165--168\relax
\mciteBstWouldAddEndPuncttrue
\mciteSetBstMidEndSepPunct{\mcitedefaultmidpunct}
{\mcitedefaultendpunct}{\mcitedefaultseppunct}\relax
\EndOfBibitem
\bibitem[Yeh \emph{et~al.}(2014)Yeh, Lain, Chiu, Liao, Moyano, Hsu, and
  Chiu]{Yeh2014}
C.-H. Yeh, Y.-W. Lain, Y.-C. Chiu, C.-H. Liao, D.~R. Moyano, S.~S.~H. Hsu and
  P.-W. Chiu, \emph{ACS nano}, 2014, \textbf{8}, 7663--7670\relax
\mciteBstWouldAddEndPuncttrue
\mciteSetBstMidEndSepPunct{\mcitedefaultmidpunct}
{\mcitedefaultendpunct}{\mcitedefaultseppunct}\relax
\EndOfBibitem
\bibitem[Sze and Ng(2006)]{Sze2006}
S.~Sze and K.~Ng, \emph{Physics of Semiconductor Devices}, Wiley, 2006\relax
\mciteBstWouldAddEndPuncttrue
\mciteSetBstMidEndSepPunct{\mcitedefaultmidpunct}
{\mcitedefaultendpunct}{\mcitedefaultseppunct}\relax
\EndOfBibitem
\bibitem[Feijoo \emph{et~al.}(2017)Feijoo, Pasadas, Iglesias, Mart{\'{\i}}n,
  Rengel, Li, Kim, Riikonen, Lipsanen, and Jim{\'{e}}nez]{Feijoo2017}
P.~C. Feijoo, F.~Pasadas, J.~M. Iglesias, M.~J. Mart{\'{\i}}n, R.~Rengel,
  C.~Li, W.~Kim, J.~Riikonen, H.~Lipsanen and D.~Jim{\'{e}}nez,
  \emph{Nanotechnology}, 2017, \textbf{28}, 485203\relax
\mciteBstWouldAddEndPuncttrue
\mciteSetBstMidEndSepPunct{\mcitedefaultmidpunct}
{\mcitedefaultendpunct}{\mcitedefaultseppunct}\relax
\EndOfBibitem
\end{mcitethebibliography}

\end{document}